\documentclass[aps,preprint,showpacs,nofootinbib]{revtex4}
\usepackage{amsfonts}
\usepackage{amsmath}
\usepackage{amssymb}
\usepackage{relsize}
\usepackage{graphicx}
\usepackage{epsfig}
\usepackage{float}
\usepackage[caption = false]{subfig}
\usepackage{color}

\usepackage{multirow}

\begin{document}

\title{Subcomponents $\psi_{4,1}^{(2d)}$ and $\psi_{3,0}^{(2c)}$ of the angular Fock coefficients: Explicit analytic representation}

\author{Evgeny Z. Liverts}
\affiliation{Racah Institute of Physics, The Hebrew University, Jerusalem 91904,
Israel}

\author{Nir Barnea}
\affiliation{Racah Institute of Physics, The Hebrew University, Jerusalem 91904,
Israel}

\begin{abstract}
The Fock expansion \cite{FOCK} describes the $S$-state wave function of the two-electron atomic system in the vicinity of the triple coalescence point.  The present work constitutes the additional appendix to our paper \cite{LEZ1} devoted to refinement and further development of calculation of the angular Fock coefficients of the Fock expansion. We derive the explicit analytic expressions for the complicated subcomponents  $\psi_{4,1}^{(2d)}$ and $\psi_{3,0}^{(2c)}$ of the angular Fock coefficients. An unusual method of using the Wolfram \emph{Mathematica} was applied.
\end{abstract}


\maketitle

\section{Introduction}\label{S0}

In our paper \cite{LEZ1} the various components of the angular Fock coefficients (AFC) were derived in the form of the one-dimensional series with fast convergence.
In particular, the solution of the individual Fock recurrence relation (see Eq.(C1))
\begin{equation}\label{C1}
\left(\Lambda^2-32\right)\psi_{4,1}^{(2d)}=h_{4,1}^{(2d)}~~~~~~~~~~~~~~~~~
\end{equation}
was represented by the subcomponent
\begin{equation}\label{718}
\psi_{4,1}^{(2d)}(\alpha,\theta)=\sum_{l=0}^\infty P_l(\cos\theta)(\sin \alpha)^l \textmd{t}_l(\alpha),
\end{equation}
where $P_l(z)$ is the Legendre polynomial. Treating $r_1,~r_2$ and $r_{12}$ as the interparticle coordinates the hyperspherical angles $\alpha$ and $\theta$ are defined as
\begin{equation}\label{1}
\alpha=2\arctan \left(r_2/r_1\right),~~~\theta=\arccos\left[(r_1^2+r_2^2-r_{12}^2)/(2r_1r_2)\right].
\end{equation}
The hyperspherical angular momentum operator reads (see Eq.(5) \cite{LEZ1})
\begin{equation}\label{2}
\Lambda^2=-\frac{4}{\sin^2 \alpha}\left(\frac{\partial}{\partial\alpha}\sin^2\alpha\frac{\partial}{\partial\alpha}+\frac{1}{\sin\theta} \frac{\partial}{\partial\theta}\sin\theta \frac{\partial}{\partial\theta}\right),
\end{equation}
and the right-hand side (rhs) of Eq.(\ref{C1}) is (see, Eqs.(C2), (C3) \cite{LEZ1})
\begin{equation}\label{C2}
h_{4,1}^{(2d)}(\alpha,\theta)=\sum_{l=0}^\infty P_l(\cos\theta)(\sin \alpha)^l \textmd{h}_l(\alpha),
\end{equation}
where
\begin{eqnarray}\label{C3}
\textmd{h}_l(\alpha)=\frac{\pi-2}{3\pi(2l-1)2^l}
\left[\sin\left(\frac{\alpha}{2}\right)+\cos\left(\frac{\alpha}{2}\right) \right]\times
\nonumber~~~~~~~~~~~~~~~~~~~~~~~~~~~~~~~~~~~~~~~~~~~~~~\\
\left[\frac{5}{(2l-3)\sin \alpha}F_{l,3}(\alpha)-\left(1-\frac{2}{\sin\alpha}\right)F_{l,1}(\alpha)-
(2l-1)F_{l,-1}(\alpha)\right]
.~~~~~~~~~~~
\end{eqnarray}
Definition
\begin{equation}\label{A8}
F_{l,\nu}(\alpha)=
~_2F_1\left(\frac{l}{2}-\frac{\nu}{4},\frac{l}{2}-\frac{\nu}{4}+\frac{1}{2};l+\frac{3}{2};\sin^2\alpha\right)
\end{equation}
was used for the Gauss hypergeometric function in the rhs of Eq.(\ref{A8}).

It was shown in \cite{LEZ1} that functions $\tau_l(\rho)\equiv t_l(\alpha)$ with $\rho=\tan(\alpha/2)$ represent the physical solution of equation
\begin{equation}\label{C5}
\left(1+\rho^2\right)^2\tau_l''(\rho)+2\rho^{-1}\left[1+\rho^2+l(1-\rho^4)\right]\tau_l'(\rho)+4(2-l)(l+4)\tau_l(\rho)=
-h_l(\rho),
\end{equation}
were (see Eq.(C6) \cite{LEZ1})
\begin{eqnarray}\label{C6}
h_l(\rho)\equiv\textmd{h}_l(\alpha)=-\frac{(\pi-2)(\rho+1)\left(\rho^2+1\right)^{l-1}}{3\pi(2l-1)(2l+3)2^{l+1}}\times
\nonumber~~~~~~~~~~~~~~~~~~~~~~~~~~~~~~~~~~~~~~~~~~~~~~~~~\\
\left[\frac{15-4l(l+1)(4l+11)}{(2l-3)(2l+5)\rho}+4l(2l+3)+2\rho+4(l+1)(2l-1)\rho^2+\frac{(2l-1)(4l+5)\rho^3}{2l+5}\right].~~~
\end{eqnarray}
 The physical solutions of Eq.(\ref{C5}) for $l=0,1,2$ were presented in \cite{LEZ1} by Eqs.(94)-(96), respectively.
Putting $A_1=0$ in Eq.(C17) \cite{LEZ1}, one obtains the physical solution of Eq.(\ref{C5}) for $l>2$ in the form
\begin{equation}\label{C17}
\tau_l(\rho)=\tau_l^{(p)}(\rho)+A_2 v_{4l}(\rho),~~~~~~~~~~~~~(l\geq 3)
\end{equation}
where the particular solution $\tau_l^{(p)}$ of the differential equation (\ref{C5}) can be obtained by Eq.(45) \cite{LEZ1} with $k=4$. This gives
\begin{equation}\label{3}
\tau_l^{(p)}(\rho)=\frac{1}{2l+1}\left[u_{4l}(\rho)\mathcal{V}_l(\rho)-v_{4l}(\rho)\mathcal{U}_l(\rho)\right],
\end{equation}
where
\begin{equation}\label{4}
\mathcal{V}_l(\rho)=\int_0^\rho v_{4l}(t)h_l(t)t^{2l+2}(1+t^2)^{-2l-3}dt,
\end{equation}
\begin{equation}\label{5}
\mathcal{U}_l(\rho)=\int_1^\rho u_{4l}(t)h_l(t)t^{2l+2}(1+t^2)^{-2l-3}dt.
\end{equation}
The solutions of the homogeneous equation associated with Eq.(\ref{C5}) is defined by Eq.(46)\cite{LEZ1} for $k=4$, which yields
\begin{subequations}\label{6}
\begin{align}
u_{4l}(\rho)=\rho^{-2l-1}(\rho^2+1)^{l+4}~_2F_1\left(\frac{7}{2},3-l;\frac{1}{2}-l;-\rho^2\right),~~~~~~~\label{6a}\\
v_{4l}(\rho)=(\rho^2+1)^{l+4}~_2F_1\left(\frac{7}{2},4+l;l+\frac{3}{2};-\rho^2\right).~~~~~~~~~~~~~~~~\label{6b}
\end{align}
\end{subequations}

\section{Derivation of the explicit representation for the particular solution $\tau_l^{(p)}$ }\label{S1}

To solve the problem mentioned in the title of this Section, it would be useful, first of all, to express the solutions $u_{4l}(\rho)$ and  $v_{4l}(\rho)$ of the homogeneous equation through the elementary functions.
It is seen from Eq.(\ref{6a}) that for $l\geq3$, we can write down
\begin{equation}\label{7}
u_{4l}(\rho)=\frac{8(l-3)!(1+\rho^2)^{l+4}}{15\sqrt{\pi}\Gamma(l+1/2)\rho^{2l+1}}
\sum_{m=0}^{l-3}\frac{(-1)^m\Gamma(m+7/2)\Gamma(l-m+1/2)}{m!(l-m-3)!}\rho^{2m}.
\end{equation}
The well-known formula (7.3.1.140)\cite{PRU3} was applied.

Solution of the problem for $v_{4l}(\rho)$, defined by Eq.(\ref{6b}), is more complicated. Making use of the relation (7.3.1.9) \cite{PRU3} for $m=l-3$ (alternatively, one can apply \cite{WS1}), and subsequent application of the transformation (7.3.1.3) \cite{WS1} (alternatively, one can apply \cite{WS2}) yields
\begin{eqnarray}\label{8}
~_2F_1\left(\frac{7}{2},l+4;l+\frac{3}{2};-\rho^2\right)=
\frac{16\Gamma(l+3/2)}{105\sqrt{\pi}\rho^{2(l-3)}(1+\rho^2)^7}
\sum_{p=0}^{l-3}\frac{(-1)^p}{p!(l-p-3)!}
\nonumber~~~~~~~~~~~~~~~~\\
\times~_2F_1\left(1,l-p+4;\frac{9}{2};\frac{\rho^2}{1+\rho^2}\right)
.~~~~~~~~
\end{eqnarray}
Next step is application of the relation (7.3.1.132) \cite{PRU3} or the Wolfram Function Site \cite{WS3} to the Gauss Hypergeometric Functions in the rhs of Eq.(\ref{8}). This gives
\begin{eqnarray}\label{9}
~_2F_1\left(1,l-p+4;\frac{9}{2};\frac{\rho^2}{1+\rho^2}\right)=
\frac{7\Gamma(l-p+1/2)}{2(l-p+3)!}
\bigg \{\frac{15(1+\rho^2)^{l-p+4}}{8\rho^8}\times
~~~~~~~~~~~~~~~\nonumber~~~~~~~~~~~~~~~~\\
\left[\frac{2\rho}{\sqrt{\pi}} \arctan (\rho)-\sum_{m=1}^3\frac{(m-1)!}{\Gamma(m+1/2)}\left(\frac{\rho^2}{1+\rho^2}\right)^m \right]+
\sum_{m=0}^{l-p-1}\frac{(l-p-m+2)!}{\Gamma(l-p-m+1/2)}(1+\rho^2)^{m+1}
\bigg\}.~~~~~~~~
\end{eqnarray}
Inserting Eqs.(\ref{8})-(\ref{9}) into Eq.(\ref{6b}), one obtains the following representation for the second solution of the homogeneous equation:
\begin{eqnarray}\label{10}
v_{4l}(\rho)=
\frac{\Gamma(l+3/2)}{\sqrt{\pi}}\left(\frac{1+\rho^2}{\rho^2}\right)^{l-3}\sum_{p=0}^{l-3}
\frac{(-1)^p \Gamma(l-p+1/2)}{p!(l-p-3)!(l-p+3)!}
\bigg \{\frac{(1+\rho^2)^{l-p+4}}{\rho^8}\times
~~~~~~~~~~~~~~~\nonumber~~~\\
\left[\frac{2\rho}{\sqrt{\pi}} \arctan (\rho)-\sum_{m=1}^3\frac{(m-1)!}{\Gamma(m+1/2)}\left(\frac{\rho^2}{1+\rho^2}\right)^m \right]+\frac{8}{15}
\sum_{m=0}^{l-p-1}\frac{(l-p-m+2)!}{\Gamma(l-p-m+1/2)}(1+\rho^2)^{m+1}
\bigg\}.~~~~~
\end{eqnarray}
To derive the analytic representation for the particular solution $\tau_l^{(p)}$, one needs to find analytic representations for the integrals (\ref{4}) and (\ref{5}) included into the rhs of Eq.(\ref{3}).
To this end, first of all, let us present the rhs of Eq.(\ref{C5}) (defined by Eq.(\ref{C6})) in the compact form
\begin{equation}\label{11}
h_{l}(t)=a_{0l}(t+1)(t^2+1)^{l-1}\sum_{n=1}^5 a_{nl}t^{n-2},
\end{equation}
where
\begin{eqnarray}\label{12}
a_{0l}=-\frac{(\pi-2)2^{-l-1}}{3\pi(2l-1)(2l+3)},~~
a_{1l}=\frac{15-4l(l+1)(4l+11)}{(2l-3)(2l+5)},~~
a_{2l}=4l(2l+3),~~
~~\nonumber~~~~~~\\
a_{3l}=2,~~
a_{4l}=4(l+1)(2l-1),~~
a_{5l}=\frac{(2l-1)(4l+5)}{2l+5}.~~~~~~~~~~~~~~~~~~~~~~~~
\end{eqnarray}
Inserting representations (\ref{7}) and (\ref{11}) into the rhs of Eq.(\ref{5}), and performing the trivial integration, one obtains
\begin{eqnarray}\label{13}
\mathcal{U}_l(\rho)=a_{0l}\frac{8(l-3)!}{15\sqrt{\pi}\Gamma(l+1/2)}
\sum_{m=0}^{l-3}\frac{\Gamma(m+7/2)\Gamma(l-m+1/2)(-1)^m}{m!(l-m-3)!}\times
~~\nonumber~~~~~~\\
\sum_{n=1}^5 a_{nl}\left(\frac{\rho^{2m+n}-1}{2m+n}+\frac{\rho^{2m+n+1}-1}{2m+n+1}\right).~~~~~~~~~~~~~~~~
\end{eqnarray}
Using Eq.(\ref{6b}) and Eqs.(\ref{11})-(\ref{12})  it is convenient to present Eq.(\ref{4}) in the form
\begin{equation}\label{14}
\mathcal{V}_l(\rho)=a_{0l}\sum_{n=0}^5 b_{n,l}\int_0^\rho t^{2l+n+1}~_2F_1\left(\frac{7}{2},l+4;l+\frac{3}{2};-t^2\right)dt,
\end{equation}
where
\begin{equation}\label{15}
b_{0,l}=a_{1l},~~b_{5,l}=a_{5l},~~b_{j,l}=a_{jl}+a_{j+1l}~~~(j=1,2,3,4).
\end{equation}
 Making use of the relation (1.16.1) \cite{PRU3} or the the Wolfram Functions Site \cite{WS4} with $a=7/2, b=l+4, c=l+3/2, \alpha=l+(p+1)/2, z=t^2$ yields
\begin{equation}\label{16}
\mathcal{V}_l(\rho)=a_{0l}\sum_{n=0}^5 \left( \frac{b_{n,l}}{2l+n+2}\right) \rho^{2l+n+2}~_3F_2\left(\frac{7}{2},l+4,l+1+\frac{n}{2};l+\frac{3}{2},l+2+\frac{n}{2};-\rho^2\right).~~
\end{equation}
The latter relation gives the representation of the integral (\ref{4}) through the generalized hypergeometric functions.

It can be useful to obtain the representation of the integral (\ref{4}) through the Gauss hypergeometric functions.
To this end, one can apply the relation (7.4.1.2) \cite{PRU3} or the Wolfram Functions Site \cite{WS5}. A simple subsequent reorganization of summation along with application of the linear transformation (7.3.1.4) \cite{PRU3} gives finally
\begin{eqnarray}\label{17}
\mathcal{V}_l(\rho)=\frac{4a_{0l}\Gamma(l+3/2)}{15\sqrt{\pi}}\rho^{2l+3}
\sum_{m=0}^2 \frac{m!}{\left(1+\rho^2\right)^{5-m}}\sum_{k=0}^m \frac{\Gamma(k+7/2)}{k!(m-k)!\Gamma(k+l+3/2)}
\left(-\frac{\rho^2}{1+\rho^2}\right)^k
\times
~~\nonumber~~~~~~\\
\left[\frac{(k+l+3)!\Gamma(l+m+3/2)}{(l+3)!\Gamma(k+l+m+5/2)}b_{2m+1,l}\rho^{2m}~
_2F_1\left(l+m-1,m-\frac{3}{2};k+l+m+\frac{5}{2};-\rho^2\right)
\right.
~~\nonumber~~~~~~\\
\left.
+\frac{b_{2(2-m),l}}{k+l-m+3}\rho^{3-2m}~_2F_1\left(l-2,m-\frac{3}{2};k+l+\frac{3}{2};-\rho^2\right)\right].~~~~~~~~~~~~~~~
\end{eqnarray}
Now we shall show that the Gauss hypergeometric functions included into Eq.(\ref{17}), in its turn, can be expressed through the elementary functions.
Use of the relation (7.3.1.10) \cite{PRU3} or the Wolfram Functions Site \cite{WS6} with $n=l-3$, and subsequent application of the linear transformation (7.3.1.3) \cite{PRU3} gives
\begin{eqnarray}\label{18}
~_2F_1\left(l-2,m-\frac{3}{2};k+l+\frac{3}{2};-\rho^2\right)=
\frac{\left(-\rho^2\right)^{3-l}\left(k+\frac{9}{2}\right)_{l-3}}{1+\rho^2}\times
~~\nonumber~~~~~~~~~~~~~~~~~~~~~~~~~~~~~~~~~~~~~~~~~~~\\
\sum_{p=0}^{l-3}\frac{(-1)^p}{p!(l-p-3)!}~_2F_1\left(1,k+p+6-m;k+\frac{9}{2};\frac{\rho^2}{1+\rho^2}\right).~~~~~~~~~~~~~~~~~~~~~~
\end{eqnarray}
In its turn, using (7.3.1.10) \cite{PRU3} or the Wolfram Functions Site \cite{WS6} with $n=l+m-2$, and then applying (7.3.1.3) \cite{PRU3}, one obtains
\begin{eqnarray}\label{19}
~_2F_1\left(l+m-1,m-\frac{3}{2};k+l+m+\frac{5}{2};-\rho^2\right)=
\frac{\left(k+\frac{9}{2}\right)_{l+m-2}}{(1+\rho^2)\left(-\rho^2\right)^{l+m-2}}\times
~~\nonumber~~~~~~~~~~~~~~~~~~~~~~~~~~~~~~\\
\sum_{p=0}^{l+m-2}\frac{(-1)^p}{p!(l+m-p-2)!}~_2F_1\left(1,k+p+6-m;k+\frac{9}{2};\frac{\rho^2}{1+\rho^2}\right).~~~~~~~~~~~~~~~~~~~~~~
\end{eqnarray}
Finally, application of the relation (7.3.1.132) \cite{PRU3} or the Wolfram Functions Site \cite{WS3} yields:
\begin{eqnarray}\label{20}
~_2F_1\left(1,k+p+6-m;k+\frac{9}{2};\frac{\rho^2}{1+\rho^2}\right)=
\frac{\Gamma(k+9/2)\Gamma(p-m+5/2)(1+\rho^2)^{k-m+p+6}}{\pi(k-m+p+5)!\rho^{2(k+4)}}\times
~~\nonumber~~~~~~~~~~~~\\
\left[ 2\rho \arctan(\rho)-\sqrt{\pi}\sum_{s=1}^{k+3}\frac{(s-1)!}{\Gamma(s+1/2)}\left(\frac{\rho^2}{1+\rho^2}\right)^s
\right]+
~~~~~~~~~~~~~~~~~~~~~~~~~~~~~~~~~~~~~~~~~~~~~~~\nonumber~~~~~~\\
\frac{2k+7}{2(k-m+p+5)}\sum_{s=0}^{p-m+1}\frac{(-1)^s\left(m-p-\frac{3}{2}\right)_s(1+\rho^2)^{s+1}}
{\left(k-m+p-s+5\right)_s},~~~~~~~~~~~~~~~~~~~~~~
\end{eqnarray}
where $(a)_n$ is the Pochhammer symbol.
Thus, Eqs.(\ref{17})-(\ref{20}) give the representation of the integral (\ref{4}) through the rational functions and the arctangent of $\rho$.
Finally, Eqs.(\ref{17})-(\ref{20}) together with Eqs.(\ref{7}), (\ref{10}) and (\ref{13}) give the partial solution $\tau_l^{(p)}$ in terms of elementary functions.

\section{Simple representation for $A_2$-coefficient}\label{S2}

In the work \cite{LEZ1} (see Appendix C) the coefficient $A_2\equiv A_2(l)$ included into the physical solution (\ref{C17}), was derived in general but very complicated (integral) form. In particular,
\begin{equation}\label{C29}
 A_2(l)=\frac{1}{\mathcal{P}_{2}(l)}\left[\frac{(\pi-2)2^{-3(l+2)}}{3\pi(2l-1)(l-2)(l+4)}\mathcal{P}_{3}(l)-\mathcal{P}_{1}(l)\right],
\end{equation}
where
\begin{equation}\label{C19}
\mathcal{P}_{1}(l)=
\int_0^{1} \tau_l^{(p)}(\rho)\frac{\rho^{2l+2}}{(1+\rho^2)^{2l+3}}d\rho,~~~~~~~~
\end{equation}
\begin{equation}\label{C20}
\mathcal{P}_{2}(l)=
\int_0^{1} v_{4l}(\rho)\frac{\rho^{2l+2}}{(1+\rho^2)^{2l+3}}d\rho=
\frac{\sqrt{\pi}~2^{-2(l+2)}\Gamma\left(l+3/2\right)}{\Gamma(l/2+3)\Gamma(l/2)},~~~~~~~~
\end{equation}
\begin{eqnarray}\label{C28}
\mathcal{P}_{3}(l)=-\frac{2^{l+1}}{(l+3)(2l-3)(2l+3)(2l+5)}\times
\nonumber~~~~~~~~~~~~~~~~~~~~~~~~~~~~~~~~~~~~~~~~~~~~~~~~~~~~~~~~~~~~~~~~~~~~~~~~~~\\
\left\{
\frac{30}{l+1}-\frac{26}{l+2}+13-4l\left[47-2l(2l(l+3)-9)\right]+2^{l+1}\times
\right.
\nonumber~~~~~~~~~~~~~~~~~~~~~~~~~~~~~~~~~~~~~~~~~~~\\
\left.
\left[
(l+1)\left[4l(l(4l+3)-17)+45\right]B_{\frac{1}{2}}\left(l+\frac{3}{2},\frac{1}{2}\right)+
8l\left[l(l(4l(l+4)+3)-56)-62\right]B_{\frac{1}{2}}\left(l+\frac{3}{2},\frac{3}{2}\right)
\right]
\right\},
\nonumber\\
\end{eqnarray}
with the incomplete beta function denoted by $B_z(a,b)$.
It is seen that even $\mathcal{P}_{3}(l)$ can not be called simple. However, the coefficient $\mathcal{P}_{1}(l)$ can be indeed called very complicated.

The \emph{Mathematica} calculation of the coefficients $A_2(l)$ for any integer $l\geq3$ shows that

1) for odd values of $l$ the coefficients $A_2(l)$ equal zero;

2) for even values of $l$ the coefficients $A_2(l)$ are reduced to the form
\begin{equation}\label{21}
 A_2(l)=\frac{2-\pi}{\pi^2}\mathcal{A}(l),
\end{equation}
with $\mathcal{A}(l)=(a_l+\pi b_l)$ where $a_l$ and $b_l$ are rational numbers.
Using the effective \emph{Mathematica} code, we have calculated the rational numbers $a_l$ and $b_l$  from $l=4$ up to $l=60$ (with step equals 2).
Making use of the \emph{Mathematica} operator \textbf{FindSequenceFunction} it is possible to find the general simple form of the coefficients $a_l$ and $b_l$.
One should emphasize, that for a given sequence there is a minimal number of terms to enable \emph{Mathematica} to find the formula of the general term.
In particular, for the coefficients $a_l$ and $b_l$ these minimal numbers are 10 and 26, corresponding to $l=4,6,8,... 22$ and $l=4,6,8,... 54$ , respectively.
Thus, application of the \emph{Mathematica} operator \textbf{FindSequenceFunction} to the sequences mentioned above yields:
\begin{eqnarray}\label{22}
\mathcal{A}(l)=\frac{\sqrt{\pi}}{360 l (l-2)\Gamma(l+\frac{1}{2})}\times
~~~~~~~~~~~~~~~~~~~~~~~~~~~~~~~~~~~~~~~~~~~~~~~~~~~~~~~~~~~~~~~~\nonumber~\\
\left\{
\frac{\left[l(l+1)(688 l^4+1376 l^3-2480 l^2-3168 l+465)+450\right]\Gamma\left(\frac{l-1}{2}\right)\Gamma\left(\frac{l+1}{2}\right)}
{(2l-3)(2l-1)(2l+1)(2l+3)(2l+5)}-\frac{56}{l-1}\left(\frac{l}{2}!\right)^2
\right\}.~~~
\end{eqnarray}
It is worth noting that inserting Eq.(\ref{3}) into the rhs of Eq.(\ref{C17}) and using (\ref{13}), one can represent the physical solution in the form
\begin{equation}\label{23}
\tau_l(\rho)=\frac{1}{2l+1}u_{4l}(\rho)\mathcal{V}_l(\rho)+\left[B_{l,0}+B_{l,1}(\rho)\right]v_{4l}(\rho),
\end{equation}
where
\begin{eqnarray}\label{24}
B_{l,0}=A_2(l)+\frac{8(l-3)!a_{0l}}{15\sqrt{\pi}(2l+1)\Gamma\left(l+\frac{1}{2}\right)}
\sum_{m=0}^{l-3}\frac{\Gamma\left(m+\frac{7}{2}\right)\Gamma\left(l-m+\frac{1}{2}\right)(-1)^m}{m!(l-m-3)!}\times
~~\nonumber~~~~~~\\
\sum_{n=1}^5 a_{nl}\left(\frac{1}{2m+n}+\frac{1}{2m+n+1}\right),~~~~~~~~~~~~~~~~
\end{eqnarray}
\begin{eqnarray}\label{25}
B_{l,1}(\rho)=-\frac{8(l-3)!a_{0l}}{15\sqrt{\pi}(2l+1)\Gamma\left(l+\frac{1}{2}\right)}
\sum_{m=0}^{l-3}\frac{\Gamma\left(m+\frac{7}{2}\right)\Gamma\left(l-m+\frac{1}{2}\right)(-1)^m}{m!(l-m-3)!}\times
~~\nonumber~~~~~~\\
\sum_{n=1}^5 a_{nl}\left(\frac{\rho^{2m+n}}{2m+n}+\frac{\rho^{2m+2n+1}}{2m+n+1}\right).~~~~~~~~~~~~~~~~
\end{eqnarray}
Given that $A_2(l)$ equals zero for odd $l$, one can derive a simple expression for $B_{l,0}$ with even $l$. A simple way to solve the problem is to make use of the method, which was applied earlier in order to find the coefficients $\mathcal{A}(l)$.
Like the latter coefficients,  $B_{l,0}$  have a form  $[(2-\pi)/\pi^2](\widetilde{a}_l+\pi \widetilde{b}_l)$, where $\widetilde{a}_l$ and $\widetilde{b}_l$ are rational numbers. Thus, calculating $\widetilde{a}_l$ and $\widetilde{b}_l$ (for even $l\geq4$) according to Eq.(\ref{24}), and then using the \emph{Mathematica} operator \textbf{FindSequenceFunction}, one finds
\begin{equation}\label{26}
B_{l,0}=\frac{2-\pi}{360 \pi^{3/2}(l-1)}\left[
\frac{\sqrt{\pi}\left(40 l^4-56 l^3+11 l^2-53 l+30\right)}{2^{l-1}l(l-2)(2l-3)(2l-1)}-
\frac{7l\left(\frac{l}{2}-2\right)!\left(\frac{l}{2}-1\right)!}{\Gamma\left(l+\frac{1}{2}\right)}
\right].
\end{equation}
The minimal length of a sequence enables the Mathematica operator \textbf{FindSequenceFunction} to find a simple function that yields the sequences $\widetilde{a}_l$ and $\widetilde{b}_l$ are 10 and 20, corresponding to $l=4,6,8,... 22$ and $l=4,6,8,... 42$, respectively.

\section{Subcomponent $\psi_{3,0}^{(2c)}$: Simple representation for coefficient $c_l$}\label{S2}

The last subcomponent we have presented in \cite{LEZ1} (see Eq.(98)) was
\begin{equation}\label{98}
\psi_{3,0}^{(2c)}(\alpha,\theta)=\sum_{l=0}^\infty P_l(\cos\theta)(\sin \alpha)^l \phi_l(\rho).
\end{equation}
It is the physical solution of the individual Fock recurrence relation (see Appendix D \cite{LEZ1})
\begin{equation}\label{D1}
\left(\Lambda^2-21\right)\psi_{3,0}^{(2c)}=h_{3,0}^{(2c)}
\end{equation}
with the rhs of the form
\begin{equation}\label{D2}
h_{3,0}^{(2c)}=\sum_{l=0}^\infty P_l(\cos \theta)(\sin \alpha)^l \textmd{h}_l(\alpha),
\end{equation}
where
\begin{equation}\label{D5}
\textmd{h}_l(\alpha)\equiv h_l(\rho)= \frac{2^{1-l}\left(\rho^2+1\right)^{l+\frac{1}{2}}\left[(1-2l)\rho^2+2l+3\right]}{3(2l-1)(2l+3)\rho}.
\end{equation}
Function $\phi_l(\rho)$ as the solution of the inhomogeneous differential equation
\begin{equation}\label{D4}
\left(1+\rho^2\right)^2\phi_l''(\rho)+2\rho^{-1}\left[1+\rho^2+l(1-\rho^4)\right]\phi_l'(\rho)+(3-2l)(7+2l)\phi_l(\rho)=-h_l(\rho)
\end{equation}
was derived in the form (see Eq.(99) \cite{LEZ1})
\begin{equation}\label{99}
\phi_l(\rho)=\phi_l^{(p)}(\rho)+c_l v_{3l}(\rho),
\end{equation}
where (see Eqs.(100) \cite{LEZ1})
\begin{equation}\label{100}
v_{3l}(\rho)=\left(\rho^2+1\right)^{l-\frac{3}{2}}
\left[\frac{(2l-3)(2l-1)}{(2l+3)(2l+5)}\rho^4+\frac{2(2l-3)}{2l+3}\rho^2+1\right].
\end{equation}
The particular solution $\phi_l^{(p)}$ of the equation (\ref{D4}) was represented in the form (see Eqs.(101)-(104) \cite{LEZ1})
\begin{equation}\label{730}
\phi_l^{(p)}(\rho)=\frac{2^{-l}\left(\rho^2+1\right)^{l-\frac{3}{2}}}{3(2l-3)(2l-1)(2l+3)(2l+5)}
\left[2f_{1l}(\rho)+\frac{2f_{2l}(\rho)+f_{3l}(\rho)}{2l+1}\right],
\end{equation}
where
\begin{equation}\label{731}
f_{1l}(\rho)=\left[9-4l(l+2)\right]\rho+\left(13-4l^2\right)\rho^3,~~~~~~~~~~~~~~~~~~~~~~~~~~~~~~~~~~~~
\end{equation}
\begin{equation}\label{732}
f_{2l}(\rho)=\left[(2l-3)(2l-1)\rho^4+2(2l-3)(2l+5)\rho^2+(2l+3)(2l+5)\right]\arctan(\rho),~~~~~~~~~
\end{equation}
\begin{eqnarray}\label{733}
f_{3l}(\rho)=-\left[(2l+3)(2l+5)\rho^4+2(2l-3)(2l+5)\rho^2+(2l-3)(2l-1)\right]\times
\nonumber~~~~~~~~~~~~~~~~~~~~\\
\frac{\rho}{l+1}~_2F_1\left(1,l+1;l+2;-\rho^2\right).~~~~~~~~~~~~~~~~~
\end{eqnarray}
It was shown (see appendix D \cite{LEZ1}) that the coefficient $c_l$ included into solution (\ref{99}) can be calculated as follows:
\begin{equation}\label{D17}
c_l=\frac{\mathcal{M}_1(l)-\mathcal{M}_2(l)}{\mathcal{M}_3(l)},
\end{equation}
where
\begin{equation}\label{D18}
\mathcal{M}_1(l)=\frac{2^{-3l-2}l!\sqrt{\pi}}{3(2l-3)(2l-1)(2l+7)\Gamma(l+3/2)}
~_3F_2\left(\frac{2l-1}{4},\frac{2l+1}{4},l+1;l+\frac{3}{2},l+\frac{3}{2};1\right),
\end{equation}
\begin{equation}\label{D19}
\mathcal{M}_3(l)\equiv\int_0^1v_{3l}(\rho)\frac{\rho^{2l+2}}{\left(\rho^2+1\right)^{2l+3}}d \rho=
\frac{2^{-l-\frac{3}{2}}(2l+1)}{(2l+3)(2l+7)}.~~~~~~~~~
\end{equation}
\begin{equation}\label{D20}
\mathcal{M}_2(l)\equiv \int_0^1\phi_{l}^{(p)}(\rho)\frac{\rho^{2l+2}}{\left(\rho^2+1\right)^{2l+3}}d \rho .
\end{equation}
It is seen that according to Eqs.(\ref{D17})-(\ref{D20}) the coefficient $c_l$ is represented by very complicated function of $l$.
However, \emph{Mathematica} calculations with any given integer $l\geq0$ show that the parameter $c_l$ has a form $c_{0,l}+c_{1,l}\pi+c_{2,l}\ln 2$ where $c_{i,l}~~(i=0,1,2)$ are rational numbers.
Using the Mathematica operator \textbf{FindSequenceFunction}, one obtains the following simple result
\begin{equation}\label{D21}
c_l=\frac{2l+1-(\pi/2)-\Phi(-1,1,l+1)}{3(2l-3)(2l-1)(2l+1)2^l},
\end{equation}
where $\Phi(z,s,a)$ is the Lerch transcendent. Note that for the case under consideration we have
\begin{equation}\label{D22}
\Phi(-1,1,l+1)=\frac{1}{2}\left[\psi\left(\frac{l}{2}+1\right)-\psi\left(\frac{l}{2}+\frac{1}{2}\right)\right]=
\frac{1}{2}\left(H_{\frac{l}{2}}-H_{\frac{l-1}{2}}\right),
\end{equation}
where $\psi(z)$ and $H_z$ are the digamma function and harmonic number, respectively.

The minimal length of a sequence enables the \emph{Mathematica} to find a simple function that yields the sequence $c_{0,l}$ is 22, whereas for $c_{1,l}$  and $c_{2,l}$  it equals 6.

\textbf{Warning}: There is the following (software) \textbf{bug} in the \emph{Mathematica} version $10.3$.

The action of the \emph{Mathematica} operators \textbf{FullSimplify} or \textbf{FunctionExpand} on the generalized hypergeometric function
\begin{equation*}
~_3F_2\left(\frac{2l-1}{4},\frac{2l+1}{4},l+1;l+\frac{3}{2},l+\frac{3}{2};1\right)
\end{equation*}
with any given integer $l$ produces a meaningless expression. One should emphasize that the previous \emph{Mathematica} versions, e.g., $10.0$ or $9.0$ give the correct results. For example, for $l=2$, one obtains:
\begin{eqnarray*}
l:=2;
~~~~~~~~~~~~~~~~~~~~~~~~~~~~~~~~~~~~~~~~~~~~~~~~~~~~~~~~~~~~~~~~~~~~~~~~~~~~~~~~~~~~~~~~~~~~~~~~~~~~~~~~~~~~~~~~~~~~~~~~~~~~~~~~\nonumber~~\\
\textbf{FunctionExpand}\left[HypergeometricPFQ\left[\left\{\frac{2l-1}{4},\frac{2l+1}{4},l+1\right\},
 \left\{l+\frac{3}{2},l+\frac{3}{2}\right\}, 1\right]\right]=
 ~~\nonumber~~~~~~~~~~~~~~~~~\\
\frac{2}{49}(-352+254\sqrt{2}).~~~~~~~~~~~~~~~~~~~~~~~~~~
\end{eqnarray*}
Notice that the considered generalized hypergeometric function is included into Eq.(\ref{D18}).

\end{document}